# Water structures reveal local hydrophobicity on the $In_2O_3(111)$ surface

Hao Chen[1,2,3,†], Matthias Blatnik[1,4], Christian L. Ritterhoff[5], Francesca Mirabella[1], Giada Franceschi[1], Michele Riva[1], Michael Schmid[1], Jan Čechal[4], Bernd Meyer[5], Ulrike Diebold[1], Margareta Wagner[1,4*]

[1] Institute of Applied Physics, TU Wien, 1040 Vienna, Austria
[2] State Key Laboratory of Catalysis, Dalian Institute of Chemical Physics, Chinese Academy of Sciences, Dalian 116023, China
[3] University of the Chinese Academy of Sciences, Beijing 100049, China
[4] Central European Institute of Technology (CEITEC), Brno University of Technology, 61200 Brno, Czech Republic
[5] Interdisciplinary Center for Molecular Materials (ICMM) and Computer Chemistry Center (CCC), Friedrich-Alexander-Universität Erlangen-Nürnberg (FAU), 91052 Erlangen, Germany
[†] present address: Materials Sciences Division, Lawrence Berkeley National Laboratory, Berkely, California 94720, US

*corresponding author: wagner@iap.tuwien.ac.at

**Abstract:** Clean oxide surfaces are generally hydrophilic. Water molecules anchor at undercoordinated surface metal atoms that act as Lewis-acid sites, and they are stabilized by H bonds to undercoordinated surface oxygens. The large unit cell of $In_2O_3(111)$ provides surface atoms in various configurations, which leads to chemical heterogeneity and a local deviation from this general rule. Experiments (TPD, XPS, ncAFM) agree quantitatively with DFT calculations and show a series of distinct phases. The first three water molecules dissociate at one specific area of the unit cell and desorb above room temperature. The next three adsorb as molecules in the adjacent region. Three more water molecules rearrange this structure and an additional nine pile up above the OH groups. Despite offering undercoordinated In and O sites, the rest of the unit cell is unfavorable for adsorption and remains water-free. The first water layer thus shows ordering into nanoscopic 3D water clusters separated by hydrophobic pockets.

Water in direct contact with inorganic materials arguably constitutes one of the most important interfaces on earth. Considerable efforts have been extended to understand how the first few water molecules interact with solid surfaces.[1] The basic tenets were established in an early review by Thiel and Madey,[2] updated by Henderson:[3] Isolated water molecules adsorb on top of metal atoms to maximize the interaction with the oxygen lone pair orbital and arrange to form H bonds with the O of neighboring water molecules. Early models for water overlayers on close-packed metal surfaces were refuted as too simplistic by more detailed investigations.[4] The direct inspection of adsorption structures with scanning tunneling microscopy (STM), backed-up by density functional theory (DFT) calculations, has shown the structural richness of adsorption configurations.[4] On Pt(111), for example, a wetting layer forms a variety of ring configurations,[5] and on the hydrophobic Cu(111) surface 3D structures evolve.[6] An even better resolution can be achieved with non-contact atomic force microscopy (AFM), which provides better resolution than STM[7] and avoids the tunneling currents that disturb fragile water arrangements.[8]

Most metals are oxidized in the ambient, and on oxides the competition to form H bonds between the O atoms of the surface and the neighboring water molecules often leads to partial water dissociation (9–11). The resulting hydroxyls are involved in the wetting of water in ambient conditions:[12,13] oxides are generally hydrophilic provided they are clean.[14,15] Some exceptions are reported, *e.g.*, a hydrophobic behavior was observed when the surface is entirely O-terminated.[16] Most oxide surfaces however offer undercoordinated surface cations (Lewis acid sites) that bind water molecules, and the interplay between H bonding to surface and water O leads to a rich structural variety observed in STM.[17]

The surface investigated here, $In_2O_3$(111), has both types of ions exposed (Figure 1a). Indium oxide crystallizes in a body-centered cubic Bravais lattice with bixbyite structure and a bulk lattice constant of 1.0117 nm. The nonpolar (111) surface exhibits a relaxed bulk termination (or unreconstructed (1×1) structure) with 3-fold symmetry, where 5- and 6-fold coordinated In(5c) and In(6c) atoms coexist with 3- and 4-fold coordinated O(3c) and O(4c). The total of 16 In atoms per unit cell (u.c.) fall into 6 categories. In Figure 1a they are labeled In-a to In-f; also shown are the four types of O(3c) atoms with labels O(α) to O(δ). The unit cell has three sites with 3-fold rotational symmetry; we call these axes and their immediate surroundings A, B, and C. The regions B at the corners of the unit cell contain four In(6c) atoms with the same coordination as in the bulk. Region A is terminated by O(3c) atoms, which are situated slightly higher than the In(5c), and C is terminated by undercoordinated In(5c). In prior work[18] we have established how to distinguish these regions with STM (Figure 1b). As is shown in the following, the large unit cell with its variety of differently coordinated surface atoms offers a regular but locally heterogeneous template for adsorbed water.

Our interest in water adsorption on indium oxide is stimulated by its traditional use as transparent conductive oxide (TCO),[19] as well as its promise as a catalyst. Recent works report efficient hydrogenation to methanol,[20,21] selective partial hydrogenation of acetylene to ethylene,[22] as well as electrocatalytic $CO_2$ reduction and photocatalytic reverse water-gas shift reactions[23,24] on various forms of indium oxide. To understand the role of this surface in de/hydrogenation reactions,[25] we have recently devised a method to probe the proton affinity of the four different types of O(3c) surface atoms.[26]

In previous works[26,27] we have found that water dissociates readily on $In_2O_3$(111) in UHV and at 300 K. The surface saturates at three dissociated water molecules/u.c. arranged symmetrically around region B (Figure 1c). Although the three dissociated water molecules are at geometrically equivalent sites, DFT calculations predicted a sequence for the binding energy per molecule of 1.28, 1.19, and 1.06 eV for adsorption of the first, second and third water molecule, respectively (see Table 1, below).[27] From DFT calculations considering up to 9 molecules/u.c., a surface phase diagram was derived for the thermodynamically most stable water adsorbate structures depending on temperature and water partial pressure.[27]

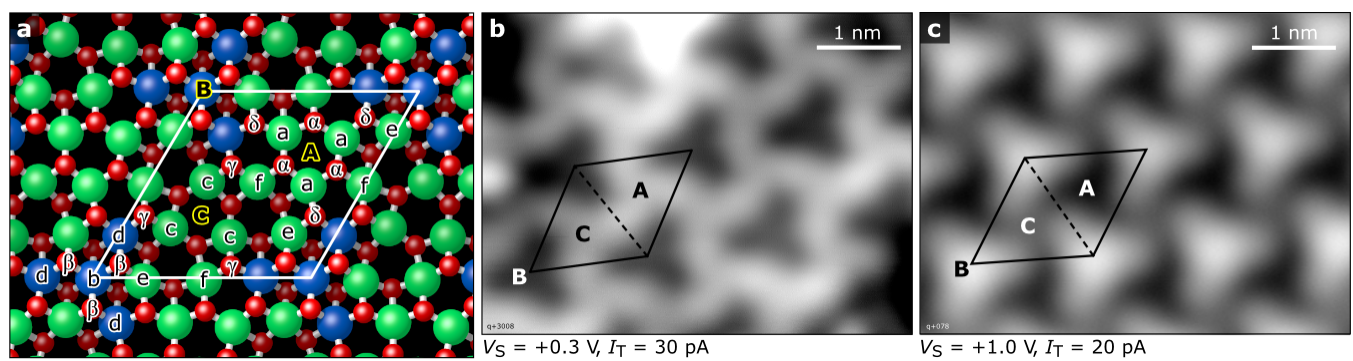

**Figure 1. The $In_2O_3$(111) surface.** (b) Atomic model highlighting the differently coordinated In (large spheres) and O (small) atoms: 5-fold coordinated In (green), 6-fold coordinated In (blue), 3-fold coordinated O (red), 4-fold coordinated O (dark red). The high-symmetry points A, B, and C are indicated. (b) STM image of the water-free surface. The In(6c) of regions B are imaged dark in empty states.[18] (c) STM image of the fully hydroxylated surface. The three OH groups and the three protonated O(3c) substrate atoms, located around region B (saturated coverage at room temperature), are imaged as a bright triangle.

The experimental results reported here confirm these predictions quantitatively. We explored a wide temperature and coverage range in UHV, and the complementary techniques of atomically resolved atomic force microscopy (AFM), temperature programmed desorption (TPD), and x-ray photoemission spectroscopy (XPS), corroborated by density functional theory (DFT), provide a full picture of water on $In_2O_3$, ranging from dissociated water molecules to multilayers. Remarkably, we find that water adsorption only occurs in regions B and C in the unit cell, whereas A stays water-free. We analyze the reason for this unusual behavior and find that it lies in the local geometry: When the O of the water molecule binds to the In-a Lewis acid site, it cannot simultaneously form a favorable H bond to an O(α). H bonding of neighboring molecules is also unfavorable because the distances are large between the In-a atoms. This leads to an exclusion of water adsorption in region A, and MD simulations of thicker water layers indicate that liquid water tends to avoid this pocket.

## Results and Discussion

**Quantitative TPD**

Figure 2 displays a series of water TPD curves obtained at beam exposures corresponding to 0–28 D$_2$O molecules/u.c. The linear relationship between the integrated area underneath the water TPD traces and the dosage (Figure 2b) indicates a constant sticking coefficient of unity for coverages up to ~40 molecules/u.c. (at 100 K). This is also in agreement with the sticking probability curves (see Supporting Information, Figure S2). Dosing labeled water, H$_2^{18}$O, shows a very similar desorption behavior with no qualitative differences in the appearance of the distinct desorption maxima.

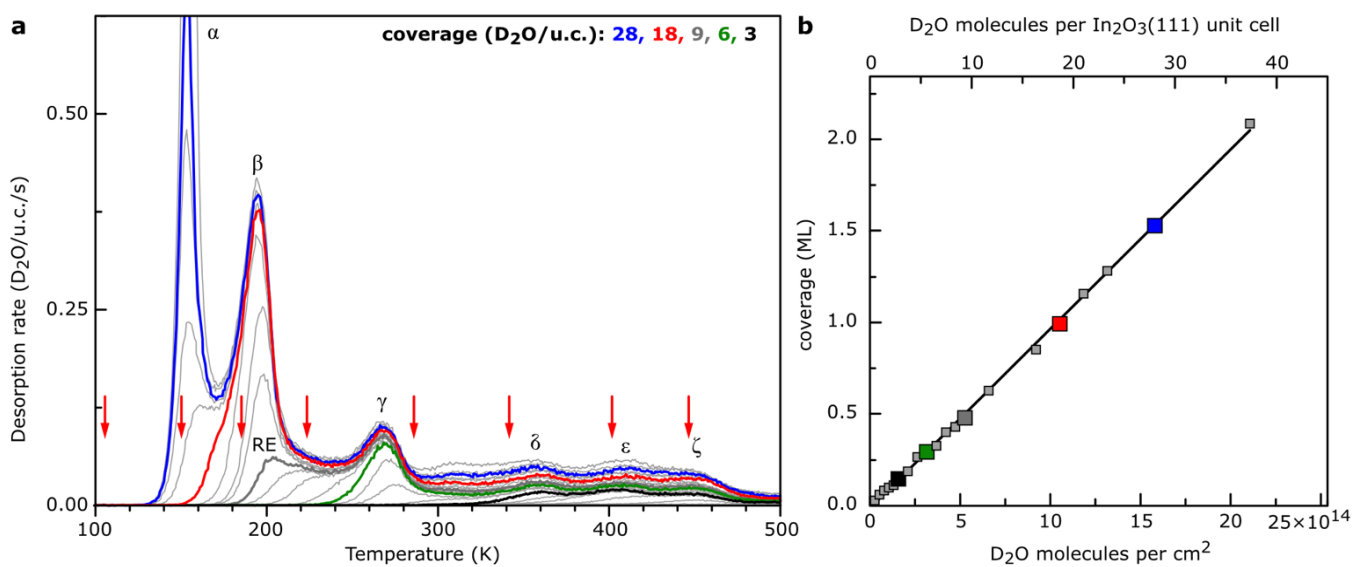

**Figure 2.** Water TPD from In$_2$O$_3$(111). (a) TPD curves of D$_2$O adsorbed at ~100 K with increasing coverage from 0 to 28 molecules/u.c. The colored curves indicate the saturated coverage for the distinct desorption peaks centered at ~152 K (α, blue), ~193 K (β, red), ~265 K (γ, green), and ~360 K (δ, black); RE indicates a reorganization phase. Red arrows point to temperatures of the XPS measurements in Figure 3. (b) Plot of the coverage (integrated area of the TPD curves; 18 molecules/u.c. are defined as 1 monolayer (ML)) as a function of exposure to the calibrated water beam. The colored squares give the coverages of the corresponding TPD traces.

Below 300 K, the TPD curves of Figure 2a have three distinct desorption peaks labeled as (with their peak maxima at) α (152 K), β (193 K), and γ (265 K). The TPD peak at the lowest temperature, α, is attributed to multilayer ice; it follows typical zero-order kinetics with a sharp leading edge (see Figure S3). Notably, the onset of this desorption peak appears at a coverage close to 18 molecules/u.c. (~$1.1\times10^{15}$ $D_2O$/cm$^2$; red curve in Figure 2a), which is slightly higher than the available 16 In surface atoms. The saturation of the β and γ peaks takes place at coverages of 18 and 6 molecules/u.c., respectively. As both peaks grow, they follow first-order desorption kinetics with their peak maxima nearly independent of the coverage. Between the β and γ peaks, a plateau develops with increasing coverage until 9 molecules/u.c., indicating a phase transformation and rearrangement of the adsorbed water species (RE). Above 300 K, three equally intense peaks develop, labeled δ (353 K), ε (402 K), and ζ (446 K). They become sequentially populated and saturate at 3, 2, and 1 molecules/u.c., respectively, in agreement with previous results that have shown a saturation of 3 hydroxyl pairs per unit cell (26, 27). Table 1 summarizes the coverages where each of these peaks reaches saturation.

| Desorption peak and $T$ | Coverage ($D_2O$/u.c.) | DFT $E_b$ (eV) | TPD $E_d$ (eV) | TPD $\nu$ (s$^{-1}$) |
|---|---|---|---|---|
| ζ (446 K) | 1 | 1.28 | 1.38±0.07 | $2\times10^{14}$ |
| ε (402 K) | 2 | 1.19 | 1.21±0.06 | $1\times10^{14}$ |
| δ (353 K) | 3 | 1.06 | 1.05±0.04 | $1\times10^{14}$ |
| γ (265 K) | 6 | 0.89 | 0.90±0.06 | $5\times10^{15}$ |
| RE (200 K) | 9 | 0.80 | 0.62±0.05 | $1\times10^{13}$ |
| β (193 K) | 18 | 0.68 | 0.61±0.04 | $5\times10^{14}$ |
| α (152 K) | 28 | 0.66 | 0.55±0.02 | - |

**Table 1.** TPD analysis and comparison to DFT. Coverages and calculated binding energies $E_b$ together with desorption energies $E_d$ and pre-factors $\nu$ of the individual TPD desorption peaks are given.

From the TPD traces, desorption energies $E_d(\theta)$ were extracted via an inversion analysis of the Polyani-Wigner equation following the procedure described in Refs. 28–30. For the first-order desorption peaks (β–ζ, RE), $\nu$ is treated as a parameter and varied from ~$10^{13}$ s$^{-1}$ to ~$10^{15}$ s$^{-1}$ (30). The highest desorption energy, ~1.38 eV, corresponds to the low-coverage limit. The lowest desorption energy, ~0.61 eV, relates to the coverage before multilayer adsorption commences; the corresponding pre-exponential factor is ~$5\times10^{14}$ s$^{-1}$. In the region of the rearranged layer (RE) the pre-exponential factor is lower, ~$1\times10^{13}$ s$^{-1}$, indicating a reordering of the adsorbed species.

**High-resolution XPS**

The chemical state of the adsorbed water species was studied with high-resolution XPS (Figure 3). This helps to assign the different peaks observed in the TPD curves to molecular and/or dissociated water and supports the coverages obtained with TPD. The water-free In$_2$O$_3$(111) surface shows a symmetric O1s peak originating from the O atoms of the In$_2$O$_3$ lattice at a binding energy of 530.3 eV (Figure 3a, acquired at 300 K). An ice multilayer was then prepared by adsorbing ~28 molecules/u.c. on the stoichiometric In$_2$O$_3$ surface at <135 K (Figure 3h). Afterward, the sample was sequentially heated at a rate of 1 K/s to specific temperatures, each one below the onset of the desorption peak of interest. The respective temperatures were kept at this value during the XPS measurements; they are indicated with red arrows in Figure 2a. The XPS spectra in Figure 3b-h correspond to TPD-derived water coverages of 1, 2, 3, 6, 9, 18, and ~28 D$_2$O/u.c. Above 450 K, all water species have desorbed from the surface except from defect sites, leading to a spectrum essentially identical to that of the clean surface. With respect to the clean surface, increasing coverage leads to a downwards band bending of up to ~0.2 eV at ≥ 9 D$_2$O/u.c.

The XPS spectra were fitted with the parameters provided in Table S1. At 405 K (ζ, 1 D$_2$O/u.c., Figure 3b), a small shoulder at the high binding energy side is resolved at 532.0 eV. We attribute this to adsorbed OH* groups from water dissociation, which agrees well with the nc-AFM results below and Ref. 27. Spectra recorded at 345 K (ε, 2 D$_2$O/u.c.) and 290 K (δ, 3 D$_2$O/u.c.) show a similar shoulder. The linear increase of the OH-related integral intensity (peak area, see Figure S4) is consistent with the coverage of dissociated water molecules, resulting in 2, 4 and 6 hydroxyl groups, respectively (see Figure 3b-d, and Figure 4).

At 227 K (γ, Figure 3e), a broader shoulder appears at higher binding energy, indicating an additional O species. The fitted peak component at 532.9 eV is attributed to molecularly adsorbed water (blue, mol-I). A comparison in integral intensity leads to a ratio of ~1:2 of mol-I:OH. Below 200 K (Figure 3f-h), as additional water is adsorbed at the surface, the O1s core level shows a new component at yet higher binding energy (534 eV; orange, mol-II), ascribed to molecular water in less direct contact with the surface. At 190 K (RE, 9 D$_2$O/u.c, Figure 3f), the ratio of the integral intensities of OH (two per dissociated water) and molecular water mol-I is ~1:1 (see Figure S4). Due to the overlap of the TPD desorption peaks β and RE (see Figure 2a), the second molecular component (534 eV) associated with the onset of the TPD β structure starts to rise. At 153 K (β, 18 D$_2$O/u.c., Figure 3g), the β structure is fully developed with a ratio of ~3:2:2 (mol-II:mol-I:OH). Below this temperature, the multilayer ice (α) condenses on the surface (Figure 3h), leading to an increase in the molecular water component at 534 eV and, eventually, attenuation of all other XPS peaks (not shown).

---

* Although the experiments were conducted with D$_2$O, we refer to the resulting hydroxyl groups as OH for simplicity.

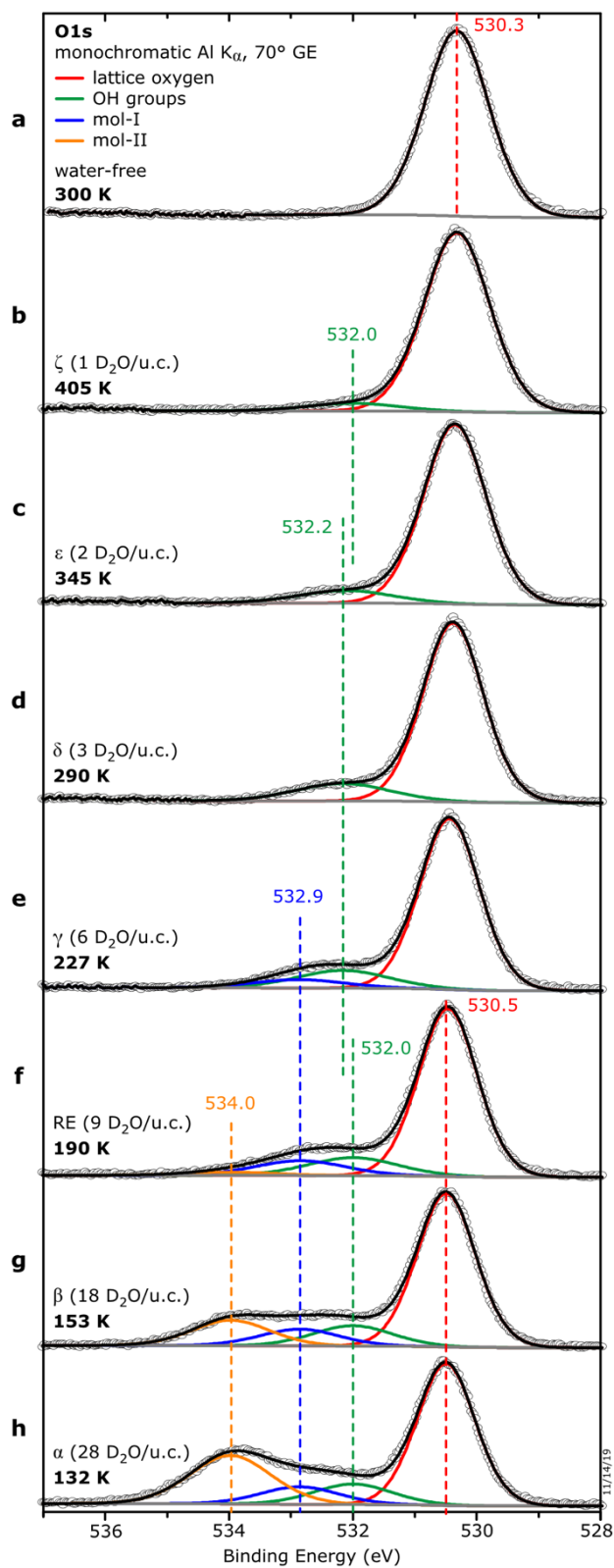

**Figure 3.** Evolution of the O1s core level upon water adsorption. The sample temperature during XPS, the corresponding TPD peaks, and the composition are, from top to bottom: (a) 300 K, bare $In_2O_3$(111) prior to water adsorption; (b) 405 K, ζ peak, one dissociated $D_2O$; (c) 345 K, ε peak, two dissociated $D_2O$; (d) 290 K, δ peak, three dissociated $D_2O$; (e) 227 K, γ peak, 3 dissociated and 3 molecular (mol-I) $D_2O$; (f) 190 K, re-arrangement, RE, 3 dissociated and 6 molecular $D_2O$ (mol-I); (g) 153 K, β peak, 3 dissociated and 15 molecular $D_2O$ (6 mol-I and 9 mol-II); (h) 132 K, α peak, 3 dissociated and ~25 molecular $D_2O$ (6 mol-I, ~19 mol-II), multilayer. The spectra were acquired at the temperatures stated, starting from the multilayer ice (panel h). The XPS peaks were fitted with hydroxyls (green), molecular water in direct (blue, mol-I) and less direct (orange, mol-II) contact with the surface; for fitting details and the peak areas as function of coverage see table S1 and Figure S4, respectively.

**Imaging with non-contact AFM and STM**

The three distinct TPD peaks above 300 K, δ, ε, and ζ (Figure 4e) correspond to the three dissociated water molecules/u.c. described previously.[27] Samples with coverages representative of these peaks were prepared by first saturating the In$_2$O$_3$(111) with water at 300 K by dosing 1 L (Langmuir, 1 L = 1×10$^{-6}$ torr·1 s) and post-annealing up to ~470 K in steps of ~20 K. After each step, the surface was inspected with STM at 80 K, see Figure S6 in the Supporting Information.

Figure 4 shows nc-AFM images obtained in a different experiment after water adsorption at 300 K. Figure 4b shows the surface after saturation, *i.e.*, at a coverage of three dissociated water molecules/u.c. Ordered features with (1×1) periodicity are observed, arranged in propeller-like structures of C$_3$ symmetry at the corner of the unit cell. Each propeller consists of three dark dots in its center, surrounded by three brighter dots (white circles). This propeller-like structure corresponds to the bright triangle observed in STM (Figure 1c and Ref. 27) and contains three pairs of OH groups arranged in the three symmetry-equivalent sites around the O(3c)/In(6c) region of the unit cell (Figure 4a). According to their adsorption sites,[26] these features are identified as O$_W$H (bright dots) and O$_S$H (dark dots), where the O atoms originate from the water molecule and the surface, respectively. The O$_W$H sticks further into the vacuum; this leads to a more repulsive interaction with the tip at constant height,[26] hence the brighter contrast. AFM images of intermediate coverages containing two (TPD peak ε) and one (TPD peak ζ) dissociated water molecules/u.c. are presented Figure 4c,d. The structures at lower coverages are fragments of the 'propeller', where one (Figure 4c) and two (Figure 4d) pairs of OH groups are missing, respectively.

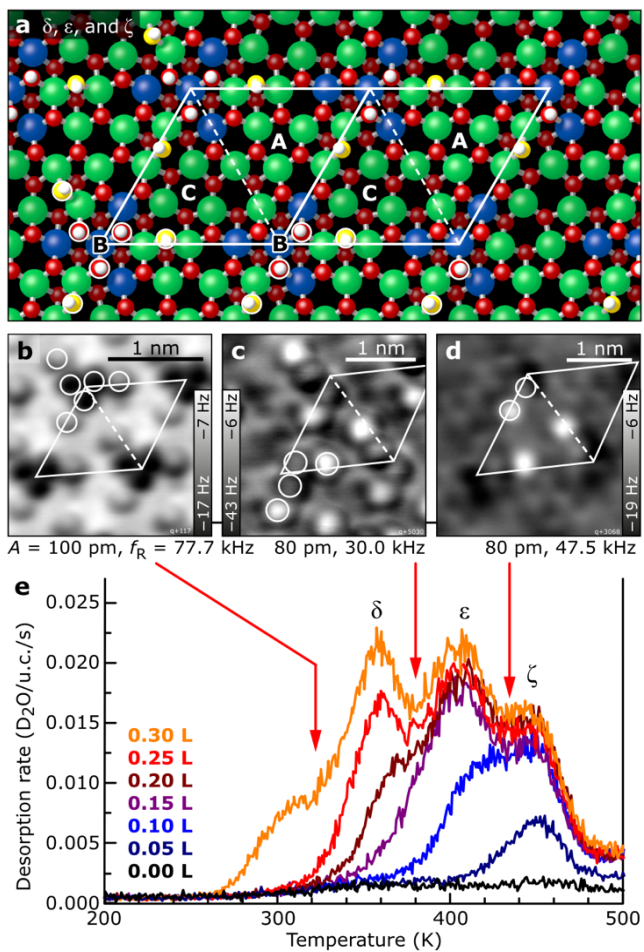

**Figure 4.** The hydroxylated $In_2O_3$(111) surface. (a) DFT-derived adsorption geometries for 1–3 dissociated water molecules/u.c. (b–d) Constant-height AFM images (5 K) of three (b), two (c), and one (d) dissociated water molecules/u.c. on $In_2O_3$(111). (e) $D_2O$ TPD curved with coverages from 0 to ~3 molecules per $In_2O_3$(111) unit cell. Three desorption peaks are centered at ~353 K (δ), ~402 K (ε), and ~446 K (ζ).

Figures 5 and 6 show AFM results of the structures below 300 K, corresponding to the desorption peaks β, RE, and γ in the TPD spectra of Figure 2a. The sample temperature during water exposure was chosen to be on the low-temperature onset of the respective desorption peak (~153 K for β, ~213 K for γ). Due to the overlapping desorption peaks of β/RE and RE/γ, combined with a different sample temperature calibration in the AFM chamber, the surfaces usually exhibited a mixture of RE and γ for the respective preparation.

In STM images, the bright triangles related to the OH groups from the three dissociated water remain visible for all structures below 300 K (shown for the β phase in Figure S7). Additional molecular water, which interacts more strongly with the STM tip, is imaged as a superimposed fuzziness (see Figure S7). The hydroxyl structure visible in the STM images was used to determine the position of the unit cell in the AFM images.

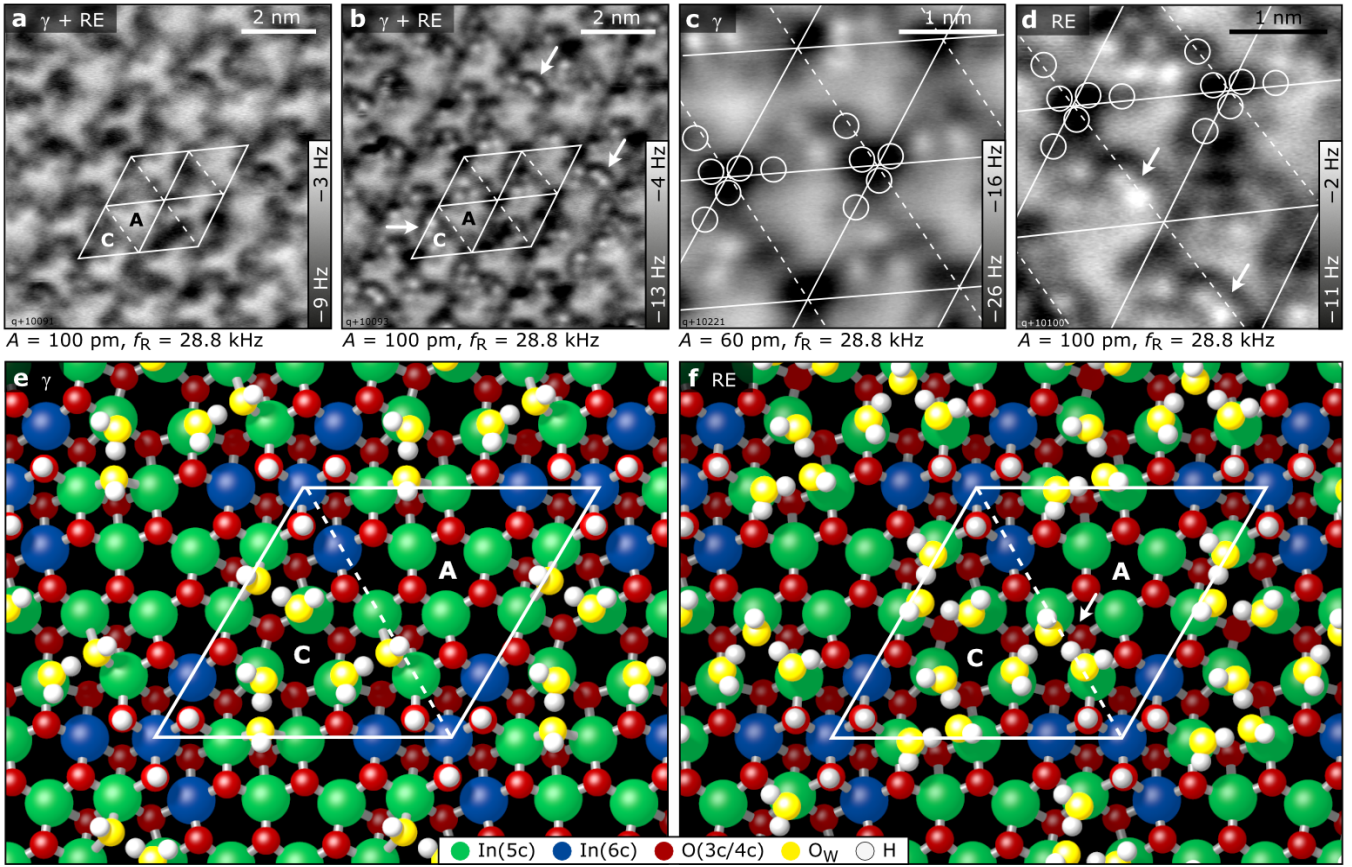

**Figure 5.** Water structures for intermediate coverages. (a, b) Constant-height AFM images taken at 5 K. Mixture of γ and rearrangement (RE) structure; the images were taken at different heights above the surface, the tip-sample separation in panel b is smaller than in panel a. Water molecules become visible as white dots in repulsive mode. Only one half of the unit cell (region C) is populated with water, while the other half (region A) remains water-free. (a, d) Higher-resolution AFM images of the γ and RE phases, respectively. The positions of hydroxyls are marked by white rings. Double features appearing in the RE phase are marked by white arrows. (e, f) DFT-derived adsorption geometries for the γ and the RE phase, respectively. (e) The γ phase contains three additional water molecules (with respect to the hydroxylated surface) in region C of the unit cell. (f) The RE phase accommodates three more $H_2O$ than γ. The $O_WH$ move and, together with the new water molecules, generate the double features in the AFM image of panel d).

In atomically-resolved AFM the contrast depends on the tip-sample separation.[7,26] Constant-height AFM images of the water structures corresponding to a mixture of the γ and RE phases are presented in Figure 5a,b. The images were acquired at different heights above the surface; Figure 5b was taken closer to the surface (higher frequency shift). The water species are visible as dark features in Figure 5a, which turn bright in Figure 5b. Notably, the additional water populates only one half of the unit cell (labeled C in Figure 1a and Figure 5), while the other half (labeled A) remains unoccupied (bright in the AFM images due to long-range interaction with the tip).

From TPD, XPS, and DFT (below), we expect the γ phase to contain three water molecules/u.c. in region C, in addition to the dissociated water propellers. In AFM, Figure 5c, these additional water molecules are imaged as bright features as they stick farther into the vacuum; the OH groups, situated lower on the surface, are thus barely visible (indicated by white circles in Figure 5c,d). The images for the γ phase show a maximum of three dots (see the unit cell in the lower half of Figure 5c), while many unit cells are occupied by only two molecules due to the overlapping desorption peaks.

When increasing the water coverage, between the TPD peaks γ and β a structural rearrangement takes place (thick grey curve in Figure 2a). According to TPD and XPS, the RE phase accommodates up to three water molecules/u.c. in addition to the γ phase. In high-resolution AFM, Figure 5d, pairs of protrusions are found at the $O_WH$ sites (white arrows). This is explained by DFT (see below): The $O_WH$, which were previously in a bridging configuration (between In-e and In-f in Figure 5e), change their adsorption site to on top and the new water molecules adsorb on the now available In sites. Thus, also the RE structure is located only in the In-terminated part of the unit cell (C in Figure 1a), while region A remains unoccupied.

The final configuration, before multilayer ice start to condense, is reflected in the β desorption peak and adds 9 water molecules to the RE structure (18 $H_2O$ in total). In the AFM images, Figure 6a,b, an incomplete array of single protrusions is visible with mostly (1×1) character. The coverage of these protrusions was below 70% in all preparations. Comparing AFM and STM images of the same surface region (Figure S7) indicates that the protrusions are located at the center of the OH propeller. The protrusions themselves are not visible in the STM images. In regions without protrusions, it is possible to go closer to the surface in constant-height AFM. Small, irregular clusters sit at the same site as the protrusions but ~2 Å closer to the surface (Figure S7). The OH groups or features of the γ and rearrangement structure are no longer visible. The coverage associated with the TPD peak β features a corrugated surface with, locally, water molecules in several layers.

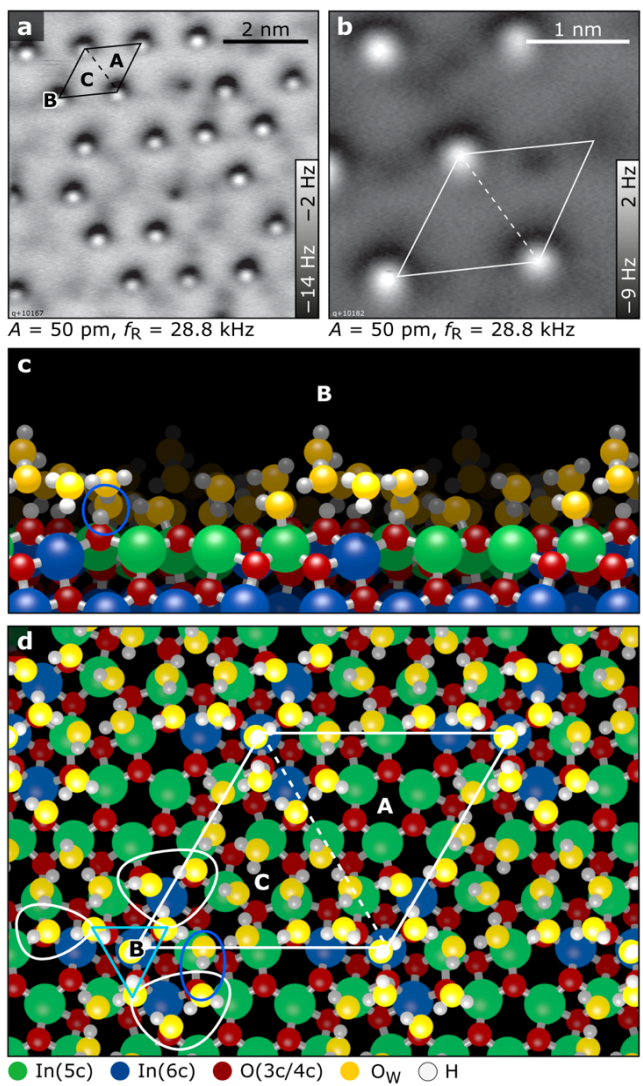

**Figure 6.** The β phase. (a, b) AFM images showing protrusions with a (1×1) periodicity. The protrusions are located in region B of the unit cell. (c, d) Side and top views of the DFT-optimized structure derived from AIMD runs. Note that region A of the unit cell is still free of water molecules.

## Computational analysis: Low and intermediate water coverage

The experimental data confirm the earlier DFT results[27] that predicted a sequence of distinct adsorbate phases of up to 9 molecules/u.c. with decreasing temperature. The XPS and AFM results of the water phases corresponding to the TPD peaks ζ, ε, δ, γ, and the RE structure (Table 1) agree with these theoretical predictions.

Above room temperature, three separate phases comprising of one, two and three water molecules/u.c. are observed. These water molecules dissociate and adsorb at symmetry-equivalent sites around B. The TPD-derived desorption energies agree with the DFT values (Table 1); the desorption energy decreases with coverage, although the resulting OH pairs occupy equivalent sites around B (Figure 4). A thorough analysis of the structural changes upon water adsorption revealed that this effect is caused by a substrate-mediated, effective repulsion between the molecules due to surface re-relaxations.[27,31] The relaxation of the surface atoms after cleavage of the crystal is partially lifted by the adsorption of the water molecules. However, because of the overlap of the spatial regions where structural changes take place, the full extent of the energy gain by this re-relaxation is only available for the first adsorbate and it is reduced for the second and third molecule.[27,31]

For higher surface coverages with up to 9 water molecules/u.c., the DFT calculations considered molecular and dissociative adsorption at all different sites of the $In_2O_3$(111) surface. Well below room temperature in UHV, two phases consisting of 6 and 9 water molecules/u.c.

were predicted.[27] Here, intact water (confirmed by XPS) adsorbs at the less preferred adsorption sites around B and C (sites In-c, In-e and In-f). The first water molecules adsorb at the three In-c sites of the unit cell while the $O_WH$ of the dissociated water molecules remain in their bridging position between In-e and In-f (γ phase with 6 molecules/u.c., Figure 5e). Increasing the coverage to 9 molecules/u.c., the $O_WH$ move to an on-top position (In-f in Figure 5f) and the intact water molecules occupy the other In-c, In-e or In-f sites (RE phase). Several alternative structures, in which the $O_WH$ groups sit on top of In-e or In-c, have similar energies, as shown by additional DFT calculations in the Supporting Information (Figure S10). These structures can easily transform into each other and some even contain a fourth dissociated water molecule.

**Computational analysis of the β phase**

At the highest coverage with 18 water molecules/u.c. in the β phase, one could expect that the remaining three In-a sites in region A become occupied by water molecules, followed by the formation of a bilayer structure. To test this assumption, additional DFT calculations with 18 adsorbed water molecules on the $In_2O_3$(111) surface were performed. While a systematic search for the global optimum structure was still possible for the adsorbate structures with up to 9 water molecules, it is no longer feasible with 18 molecules. Thus, we turned to *ab initio* molecular dynamics (AIMD) simulations[32] for identifying important representative structural motifs within the high-coverage β phase.

An ensemble of 24 structures was created by a simulated-annealing procedure. Snapshots were selected from an 80 ps AIMD simulation at 360 K and were quenched to zero temperature at different rates (see Supporting Information). The 12 energetically most favorable structures from this ensemble are within an energy range of 0.1 eV. The structure with the lowest energy is shown in Figure 6. Some other representative low-energy structures of the simulated annealing search are shown in Figure S8. Although all configurations are slightly different, they have several characteristic features in common (with some minor exceptions, see Supporting Information):

Nine out of the 18 water molecules are adsorbed at In(5c) surface sites and form a first water layer. Three of these water molecules remain dissociated, with the protons on the O(β) next to B and the $O_WH$ at In(5c) in close vicinity, in an arrangement reminiscent of the low-coverage dissociative adsorption. Often a fourth dissociated molecule is found (dark blue oval in Figure 6c,d). The proton sits preferentially on an O(δ), which is the site with the second highest proton affinity next to the already protonated O(β).[26] Most importantly, the four $O_WH$ groups of the dissociated water molecules together with the five remaining intact molecules are almost exclusively found on the In(5c) sites c, e, and f next to B and C. The In-a in region A remain almost always water-free! Only in a few structures of our ensemble one out of the 9 molecules of the second layer has moved to an In-a site. Altogether, the water molecules in the first water layer form a very similar structure as in the RE phase.

The remaining 9 of the 18 water molecules are located above region B and C and form a second water layer. They are connected to the surface only by H bonds to surface O(δ) and O(γ) atoms, or to one of the other 9 water molecules in the first layer, in good correspondence with the peak fit in XPS (Figure 3, mol-II, orange). Only in about one third of the structures of our ensemble a water molecule from the second layer is connected to one of the three O(α) by a H bond. This is a further indication of the hydrophobic nature of region A.

Four of the 9 water molecules in the second layer always form a triangular 'cap' above B (see light blue triangle in Figure 6d): three molecules receive an H bond from the protonated surface O(β), the fourth one is sitting in the center of the triangle of the O(β) directly above In-b. In 10 out of the 12 lowest-energy structures of our simulated annealing search, one of the four water molecules is oriented such that an OH end sticks perpendicularly out of the second layer, preferentially from the water molecule sitting in the center of the triangle, see Figure 6c,d. This OH group is responsible for the pronounced contrast in the AFM images, see Figure 6a,b. However, the cap of 4 water molecules alone is not stable at this position. In test calculations, where such a cap was added on top of the first water layer of 9 molecules, the cap relaxed away from the center of the high-symmetry site B. Additional water molecules at the periphery are required to stabilize the triangular cap at B (see the three white circles in Figure 6d). There are always two more water molecules at the corner that is closest to the fourth dissociated water molecule in the first layer (see white circle in Figure 6d). At the other two corners there are one or two more molecules. Thus, the peripheral ring consists of 4 to 6 water molecules with an average of 5. They saturate the H bonds from the central molecules of the cap and connect to the surface by donating H bonds to surface O atoms. Thus, on average 9 water molecules are required to form a stable, low-energy second-layer structure, with variation between 8 and 10. The stability of the structure is reflected by the calculated average binding energy of 0.68 eV for the 9 molecules of the second layer, which is larger than the binding energy of a water molecule in ice (0.66 eV). This explains why the β structure gives rise to a distinct desorption peak in TPD at a temperature above multilayer desorption. We note that the experimentally derived desorption energy of the β peak agrees with the one derived from DFT for this structure (Table 1). Also, the XPS spectra (Figure 3) show distinct features due to molecular water, again in agreement with the various water configurations described here.

**Why is region A hydrophobic?**

It is remarkable that additional water molecules, beyond a coverage of 9 molecules/u.c., prefer to form a cluster above B instead of adsorbing at the undercoordinated In(5c) and O(3c) sites in region A. The interaction of the water oxygen lone pair orbital with an In(5c) and the formation of an H bond to an O(3c) usually leads to a rather strong adsorption of a water molecule on such pairs of Lewis acid and base surface sites. However, additional DFT calculations show that the binding energy of water molecules in region A stays below 0.57 eV, independent of the water pre-coverage of the surface (Table S2 and Figure S11). This is smaller than the calculated binding energy of water molecules in ice (0.66 eV) or in the water cluster above B (0.68 eV).

There are two main reasons for this result: First, the lone pair interaction with the In(5c) site and the H bond to O(3c) are very directional. Ideally, the O atom of the water molecule should occupy the position of an O lattice atom above the In(5c), and its H bond should be oriented along the direction to an In lattice position above a neighboring O(3c). On $In_2O_3$(111), the In-a and O($\alpha$) form a slightly puckered, almost planar 6-membered ring. Also, the neighboring O($\delta$) are basically within the same plane (see Figure S12). In this geometry, a water molecule cannot establish simultaneously both directional interactions with the Lewis acid (the In) and Lewis base (the O) surface sites. It is instructive to compare this situation with the adsorption of single water molecules on another post-transition metal oxide, the $ZnO(10\bar{1}0)$ surface,[9,33] see Figure S12. Here, a large water binding energy of 0.94 eV is found for a geometry in which the water molecule sits across a trench of the $ZnO(10\bar{1}0)$ surface, thereby fulfilling almost perfectly both geometric requirements for a strong Lewis acid/base interaction. However, for water molecules on top of a surface ZnO dimer, the binding geometry is similar to that in region A on $In_2O_3$(111): With a frustrated Lewis acid/base interaction, the binding energy drops to 0.57 eV,[33] the same value as for adsorption on In-a sites of the $In_2O_3$(111) surface (Figure S12).

The second reason for the low water binding energy at A is that due to the almost planar geometry of the In(5c) and O(3c) sites around A, the distance between water molecules, pinned by the directional lone pair interaction to the on top position above the In(5c), is too large for the formation of strong H bonds between them. This is the case for pairs and trimers of water molecules on In-a of the same A site, and it is also true between water molecules adsorbed at In-a and other, neighboring In sites, see the Supporting Information.

Consistent with our experimental observations, the small pocket in region A remains uncovered when water is frozen onto this surface up to a coverage of 18 H$_2$O/u.c. Preliminary MD calculations of thicker water films (Figure S13) show that this area of the unit cell also remains inhospitable for water molecules in thicker water films. Since neither a single water molecule nor a water cluster will find a favorable adsorption configuration, the residence probability in this part of the unit cell is reduced, pointing to a locally hydrophobic behavior on an otherwise hydrophilic surface.

## Summary and Conclusions

This joint theoretical and experimental study provides a complete and consistent description of water adsorption on a post-transition metal oxide surface, In$_2$O$_3$(111), that is of increasing interest in heterogeneous catalysis. The results confirm and directly demonstrate many well-established concepts that govern oxide–water interactions, while other, at first sight surprising observations can be explained satisfactorily (or have even been predicted[27]) by analysis based on detailed calculations. The part of the In$_2$O$_3$(111) unit cell that contains undercoordinated O atoms with the highest proton affinity[26] actively dissociates water molecules; the difference of 0.1 eV in binding energy of these otherwise completely equivalent hydroxyl pairs results in a ~50 K variation of their desorption temperature. The 'propeller' of hydroxyls, clearly imaged with non-contact AFM, acts as a nucleation point for the build-up of nanoscopic 3D water clusters. In their immediate neighborhood, a flat arrangement of Lewis acid/base sites provides an unfavorable adsorption template. This nanoscopic checkerboard suggest an overall amphiphilic nature of this oxide surface: Within the nanometer-sized unit cell, regions B and A attract and repel water molecules, respectively, suggesting that region A is not blocked by water and remains accessible to non-polar molecules in the aqueous phase. So-called 'hydrophobic pockets' provide much of the functionality of biomolecules. This is unusual for an inorganic surface and might be useful for tailoring surface reactions.

## Methods

**Experimental Methods**

The experiments were conducted in two separate UHV chambers. The home-built molecular-beam-based TPD system, described in more detail in Ref. 34, has a base pressure of ~1×10$^{−10}$ mbar and is equipped with a liquid-helium flow cryostat (sample base temperature

~50 K), a gas-dosing system with a molecular beam with a diameter of 3.1 mm at the sample surface, a mass spectrometer (Hiden Analytical) for high sensitivity TPD measurements, and XPS using a monochromatic Al K$_\alpha$ source (1486.6 eV, SPECS Focus 500) with a SPECS Phoibos 150 hemispherical analyzer. All XPS measurements were performed in grazing emission (~70° with respect to the surface normal).

The AFM images were obtained in a different UHV system with an Omicron LT-AFM at 5 K employing qPlus sensors.[7] The sensors used had the following characteristics: (1) $f_R$ ≈ 30 kHz, $k$ = 2000 N/m, $Q$ ≈ 36,000, (2) $f_R$ = 47.5 kHz, $k$ = 3750 N/m, $Q$ ≈ 15,000, (3) $f_R$ = 77.7 kHz, $k$ = 5400 N/m, $Q$ ≈ 71,000, and (4) $f_R$ = 28.8 kHz, $k$ = 2000 N/m, $Q$ ≈ 45,000. The tips (etched W) were prepared on a Cu crystal and also on the $In_2O_3$(111) surface by voltage pulses until the frequency shift at +1 V sample bias and 20 pA tunneling current was less then −1 Hz and −5 Hz on Cu and $In_2O_3$, respectively. An oscillation amplitude of 80–130 pm was used for constant-height imaging; the amplitudes together with the $f_R$ are indicated at each image or in the figure caption. Small amounts of water were dosed via a leak valve admitting ~1×10$^{-9}$ mbar.

For the combined TDP/XPS study 190 nm thick $In_2O_3$(111) films grown by pulsed laser deposition on 5×5 mm$^2$ YSZ substrates were used.[35] The surface properties of the $In_2O_3$(111) films and the $In_2O_3$(111) single crystals are equivalent as confirmed by STM. The nc-AFM measurements were performed on $In_2O_3$(111) single crystals. The surfaces of both types of samples were initially cleaned by cycles of sputtering (1 keV Ar$^+$ ions, normal incidence) and annealing in oxygen (~1×10$^{-7}$ mbar, ~750 K); between the water experiments re-oxidation and/or sputtering and annealing was employed.

In all TPD experiments, specific amounts of heavy water, $D_2O$, were dosed directly onto the sample, which was kept at ~100 K and subsequently heated up to 500 K for each TPD curve, with a rate of 1 K/s. The TPD experiments were carried out in random order to exclude any memory effect. From prior work it is known that surface reduction and the formation of In adatoms starts below 570 K.[18] To ensure the surface stayed unchanged, the TPD heating ramps were stopped at 500 K. As a consequence, desorption originating from defects is not completely covered by the temperature range, visible in the nonzero intensity at 500 K. The desorption traces of mass-to-charge ratio m/z = 20 are displayed in Figure 2 and Figure 4. In addition, the m/z ratios 18 ($H_2O^+$ and $DO^+$), 17 ($OH^+$), 28 ($CO^+$), and 44 ($CO_2^+$) were monitored. Due to the utilization of labeled water the desorption of $D_2O$ is disentangled from co-adsorbing $H_2O$ from the residual vacuum. The measured water signal corresponds directly to the water deposited onto the $In_2O_3$(111) surface. The calibrated molecular beam allows exposing the sample to a controlled and precise amount of water by adjusting the beam flux (1.05×10$^{13}$ molecules cm$^{-2}$·s$^{-1}$ in our experiments) and exposure time.

Under the assumption of first order desorption kinetics and a coverage-dependent desorption energy ($E_d$), the desorption process can be described by the Polanyi-Wigner equation

$$-\frac{d\theta}{dt} = -\beta\frac{d\theta}{dT} = \nu\theta\exp\left(-\frac{E_d(\theta)}{k_b T}\right)$$

where θ is the adsorbate coverage, β is the heating rate, and ν is the pre-exponential factor. Through an inversion of this formula, the value of $E_d(\theta)$ can be extracted. Here, ν is treated as a parameter and varied from ~10$^{13}$ s$^{-1}$ to ~10$^{15}$ s$^{-1}$. By following the procedure described in Ref. 28–30, *i.e.*, by fitting the TPD series to find the best ν, the desorption energies were determined (see Table 1).

**Theoretical Methods**

Structure relaxations of water molecules on In$_2$O$_3$(111) were carried out with the periodic plane-wave DFT code `PWscf` of the Quantum Espresso software package,[36] using the PBE exchange-correlation functional of Perdew, Burke, and Ernzerhof,[37] Vanderbilt ultrasoft pseudopotentials[38] and a plane-wave basis set with an energy cutoff of 30 Ry. The In$_2$O$_3$(111) surface was represented by a periodically repeated slab with a thickness of four O$_{12}$–In$_{16}$–O$_{12}$ triple layers and a primitive (1×1) surface unit cell (160 atoms). The PBE-optimized bulk lattice constant of 10.276 Å was used for the lateral slab dimensions. The atoms in the two bottom layers of the slab were kept frozen in their bulk positions and only the upper layers and the adsorbed water molecules were allowed to relax. The force convergence threshold was set to 5 meV/Å. A (2,2,1) Monkhorst-Pack k-point mesh for Brillouin zone integrations was sufficient for obtaining well-converged structures and binding energies. All reported water binding energies, calculated as the total energy difference between the relaxed fragments and the slab with adsorbed water molecules, are given without corrections for zero-point vibrational energies (ZPVE) and finite-temperature contributions.

The reference calculation for bulk ice was done using the proton-ordered Ih model of Bernal and Fowler.[39,40] The hexagonal unit cell with space group symmetry P6$_3$cm contains 12 water molecules. With the PBE functional we obtain a value of 0.66 eV for the lattice energy, which is defined as the sublimation energy of a water molecule without corrections for the ZPVE and the quantum nature of the proton (see Table 1), in good agreement with previous PBE calculations.[41] To compare this result with the experimental binding energy of water multilayers in the α phase, we determined the ZPVE correction to the lattice energy by calculation of vibrational frequencies using finite differences. We find that the ZPVE lowers the formation energy by 0.119 eV. This is identical to the value derived from experimental IR spectra.[42] For $D_2O$, which was used in the experiments, the calculated ZPVE correction reduces to 0.094 eV. Thus, the ZPE-corrected theoretical value of 0.57 eV for the binding energy of water molecules in ice is in very good agreement with the experimental result of 0.55 eV for the α phase, see Table 1 in the manuscript.

The *ab initio* molecular dynamics (AIMD) simulations were performed with the Car-Parrinello Molecular Dynamics (CPMD) code[43] using the version with our recent code optimizations.[32] All settings concerning the functional, pseudopotentials, plane-wave basis set and $In_2O_3$(111) slab were kept identical as in the `PWscf` geometry optimizations. A time step of 6 a.u. (0.145 fs) was used for the integration of the equations of motion and the fictitious electronic mass was set to 700 a.u. All hydrogen atoms were replaced by deuterium.

**Funding Sources**

This work was supported by the Austrian Science Fund (FWF), project V 773-N (Elise-Richter-Stelle, M. Wagner), the European Research Council (ERC) under the European Union's Horizon 2020 research and innovation program (grant agreement No. 883395, Advanced Research Grant 'WatFun', U. Diebold and G. Francesci; grant agreement 810626, project 'SINNCE', M. Wagner, M. Blatnik, and J. Cechal), and the German Research Foundation (DFG), Research Unit FOR 1878 ('funCOS', B. Meyer).

**Acknowledgment**

H. Chen acknowledges the 'Joint PhD Training Program' awarded by the University of the Chinese Academy of Sciences (UCAS). Computational resources were provided by the Erlangen National High Performance Computing Center (NHR@FAU).

**Supporting Information Available**: Additional Experimental Data: Temperature Programmed Desorption; TPD: Sticking Probability Curve of Water on $In_2O_3$(111); TPD: Multilayer Desorption; XPS Peak Fitting; STM: Water Coverage as Function of Temperature; AFM: The β Phase. Additional Theoretical Calculations: Structures with 18 Adsorbed Water Molecules (β Phase); Structures with 9 Adsorbed Water Molecules (RE Phase); Water Adsorption in Region; MD Simulation for a Thick Water Film on $In_2O_3$(111).

**Competing interests:** The authors declare that they have no competing interests.

**TOC image**

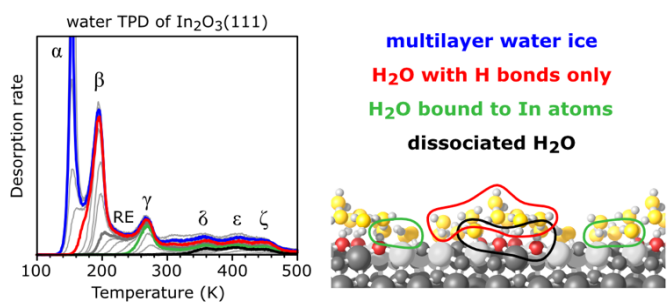